# Optimal Traffic Organisation in Ants under Crowded Conditions.

Audrey Dussutour[*‡], Vincent Fourcassié[*], Dirk Helbing[#], Jean-Louis Deneubourg[‡]

[*]*Centre de Recherches sur la Cognition Animale, UMR CNRS 5169, Université Paul Sabatier, 118 Route de Narbonne, F-31062 Toulouse Cedex 4, France*

[‡]*Centre d'Études des Phénomènes Non-linéaires et des Systèmes Complexe, Université Libre de Bruxelles, CP231, Boulevard du Triomphe, 1050 Bruxelles, Belgium*

[#]*Institute for Economics and Traffic, Dresden University of Technology, D-01062 Dresden, Germany*

**Efficient transportation, a hot topic in nonlinear science[1], is most essential for modern societies and the survival of biological species. As biological evolution has generated a rich variety of successful solutions[2], nature can inspire optimisation strategies[3,4]. Foraging ants, for example, form attractive trails which support the exploitation of initially unknown food sources in almost the minimum possible time[5,6]. However, can this strategy cope with bottleneck situations, when interactions cause delays, which reduce the overall flow? Here, we present an experimental study of ants confronted with two alternative ways. We find that pheromone-based attraction generates one trail at low densities, while at a high level of crowding, another trail is established before the traffic volume is affected, which guarantees that an optimal rate of food return is maintained. This bifurcation phenomenon is explained by a non-linear modelling approach. Surprisingly, the underlying mechanism is based on inhibitory interactions. It implies capacity reserves, a limitation of the density-induced speed reduction, and a sufficient pheromone concentration for reliable trail perception. The**

**balancing mechanism between cohesive and dispersive forces appears to be generic in natural, urban, and transportation systems.**

Animals living in groups[7,8] often display collective movement along well-defined lanes or trails[9-13]. This behaviour emerges through self-organisation resulting from the action of individuals on the environment[14-16]. In ants[17,18] for example, mass recruitment allows a colony to make adaptive choices solely based on information collected at the local level by individual workers. When a scout ant discovers a food source, it lays an odour trail on its way back to the nest. Recruited ants use the trail to find the food source and, when coming back to the nest, reinforce it in turn. Without significant crowding, mass recruitment generally leads to the use of only one branch, i.e. to asymmetrical traffic[19], because small initial differences in pheromone concentration between trails are amplified. This also explains why ants use the shorter of two branches leading to the same food source[20], if it does not constitute a bottleneck.

Here, we investigate the regulation of traffic flow during mass recruitment in the black garden ant *Lasius niger*. Ants were forced to move on a diamond-shaped bridge with two branches between their nest and a food source (Fig. 1). We have studied to what extent the traffic on the bridge remains asymmetrical and, therefore, limited by the capacity of one branch, when an increased level of crowding is induced by using branches of reduced widths ($w$ = 10.0, 6.0, 3.0 and 1.5 mm). The temporal evolution of the flow of ants on the bridge shown in Fig. 2 is typical of a trail recruitment process[21]. Surprisingly, the recruitment dynamics was not influenced by the branch width $w$ and the traffic volumes were the same (ANOVA, $F_{3,44}$ = 0.500, $P$ = 0.690). However, for $w$=10 mm, the majority of ants used the same branch (Fig. 3a), while for $w \leq 6$ mm most experiments led to symmetrical traffic ($\chi_2$ = 1.686, d.f. = 2, $P$ = 0.430). This suggests that there is a transition from asymmetrical to symmetrical traffic between 10.0 and 6.0 mm ($\chi_2$= 12.667, d.f. = 3, $P$ = 0.005).

The mechanism for this "symmetry restoring transition" has been identified by experiments, analytical calculations, and Monte Carlo simulations. Figure 3b supports indeed that the proportion of experiments producing symmetrical flows on narrow bridges increases with the total number of ants crossing the bridge ($\chi_2 = 4.6$, d.f. = 2, $P = 0.04$). Moreover, Fig. 3c shows that both branches were equally used by the opposite flow directions. Therefore, in contrast to army ants[6] and pedestrians[1], separation of the opposite flow directions was not found. This could be explained by a high level of disturbance, e.g. a large variation in the speeds of ants[22].

The flow of ants over the bridge can be understood analytically: The concentration of the trail pheromone $C_{ij}$ on branch $i$ ($i = 1, 2$) immediately behind each choice point $j$ ($j = 1, 2$) changes in time $t$ according to

$$dC_{ij}/dt = q\Phi_{ij}(t) + q\Phi_{ij'}(t-\tau) - \nu C_{ij}(t) \quad \text{with} \quad j' = 3 - j, \qquad (1)$$

where $\Phi_{i1}(t)$ represents the overall flow of foragers from the nest to the food source choosing branch $i$ behind the choice point 1, $\Phi_{i2}(t)$ the opposite flow on branch $i$ behind the other choice point $j' = 3 - j = 2$, $\tau$ the average time required for an ant to get from one choice point to the other, $q$ the quantity of pheromone laid on the trail, and $\nu$ the decay rate of the pheromone. Moreover, if the density is low enough,

$$\Phi_{ij}(t) = \phi_j(t) F_{ij}(t), \qquad (2)$$

where $\phi_1$ is the outbound flow of foragers from the nest to the food source and $\phi_2$ the opposite, nestbound flow. The function $F_{ij}$ describes the relative attractiveness of the trail on branch $i$ at choice point $j$. For *Lasius niger* it is given by[19]

$$F_{ij} = \frac{(k+C_{ij})^2}{(k+C_{1j})^2 + (k+C_{2j})^2} = 1 - F_{i'j} \quad \text{with} \quad i' = 3-i, \tag{3}$$

where $k$ denotes a concentration threshold beyond which the pheromone-based choice of a trail begins to be effective.

Equation (2) describes the flow dynamics without interactions between ants. However, when traffic was dense, we observed that ants that had just engaged on a branch were often pushed to the other branch when colliding frontally with another ant coming from the opposite direction. This behaviour will turn out to be essential in generating symmetrical traffic on narrow bridges. While pushing was practically never observed on a 10 mm bridge, on narrow bridges the frequency of pushing events was high and proportional to the flow (Fig. 3d).

When pushing is taken into account, the overall flow of ants arriving at choice point $j$ and choosing branch $i$ can be expressed by the following formula:

$$\Phi_{ij}(t) = \phi_j(t) F_{ij}(t) [1 - \gamma a \Phi_{i'j}(t-\tau)/w] + \phi_j(t) F_{i'j}(t) \gamma a \Phi_{i'j'}(t-\tau)/w \tag{4}$$

The first term on the right-hand side of (4) represents the flow of ants engaged on branch $i$ $[\phi_j(t) F_{ij}(t)]$, diminished by the flow of ants pushed towards the other branch $i'$ by ants arriving from the opposite direction. $a\Phi_{ij'}(t-\tau)/w$ is the proportion of ants decelerated by interactions. The factor $a$ is proportional to the interaction time period and the lateral width of ants. Up to the symmetry restoring transition it can be considered approximately constant. $\gamma \approx 0.57$ denotes the probability of being pushed in case of an encounter (Fig. 3d). The second term on the right-hand side of Eq. (4) represents the flow of ants that were engaged on branch $i'$ and were pushed towards branch $i$.

The stationary solutions of this model are defined by the conditions $dC_{ij}/dt = 0$, $F_{ij}(t-\tau) = F_{ij}(t) = F_{ij}$, $\Phi_{ij}(t-\tau) = \Phi_{ij}(t) = \Phi_{ij}$, and $\phi_j(t-\tau) = \phi_j(t) = \phi_{j'}(t) = \phi$ (as the nestbound flow and the outbound flow should be equal). These imply

$C_{ij} = q(\Phi_{ij} + \Phi_{ij'})/\nu = C_{ij'}$, $F_{ij} = F_{ij'} = 1 - F_{i'j} = 1 - F_{i'j'}$

1 and $C_{ij} = C_{ij'} = \frac{q\phi}{v} + D$, $\quad C_{i'j} = C_{i'j'} = \frac{q\phi}{v} - D$ (5)

2 with $\quad D = 0$ (6)

3 or (7)
$$D^2 = \frac{q^2\phi^2}{v^2} - \frac{k^2 + \gamma a\phi(k + 2q\phi/v)^2/w}{1 + \gamma a\phi/w}$$

4 We can distinguish two cases: i) When $D^2 > 0$, the stable stationary solution corresponds to
5 asymmetrical traffic with $C_{ij} = \frac{q\phi}{v} \pm \sqrt{D^2}$ and $C_{i'j} = \frac{q\phi}{v} \mp \sqrt{D^2}$. If $\gamma = 0$, this situation is
6 given for $q\phi/v > k$. If $q\phi/v \leq k$, i.e. if the pheromone concentration is too low, the ants
7 choose both branches at random, corresponding to a symmetrical distribution (Fig. 4a). ii)
8 When pushing is taken into account with $\gamma > 0$, the organization of traffic changes
9 considerably: the asymmetrical solution cannot be established anymore as soon as $D^2 < 0$ for
10 large traffic volumes $\phi$ (Fig. 4b). In this case, symmetrical traffic is expected with
11 $C_{ij} = q\phi/v$ and $\Phi_{ij} = \phi/2$ for both branches *i* and choice points *j*. Therefore, high traffic
12 volumes can be maintained and none of the branches is preferred despite of the competitive
13 effect due to the accumulation of pheromone on both branches. Moreover, the model implies
14 that the outbound flow $\Phi_{i1}$ and the nestbound flow $\Phi_{i2}$ are equal, indicating that one-way
15 flows are not required to maintain a high traffic volume[23]. These analytical results have been
16 confirmed by Monte-Carlo simulations (see Fig. 4c,d). Our simulations have also
17 demonstrated that, if pushed ants, instead of moving to the other branch, made a U-turn in the
18 direction they were coming from, the overall flow of ants crossing the bridge was affected by
19 the branch width and no shift to symmetrical traffic was observed.

20 The traffic organisation in ants can be called optimal. The overall flow on branch *i* behind
21 choice point *j* is given by $\Phi_{ij} = w\rho_{ij}V_{ij} \leq \phi$, where $\rho_{ij}$ denotes the density of ants. Their
22 average speed is theoretically estimated by $V_{ij} \approx V_m(1 - a\Phi_{ij'}/w)$, where $V_m$ denotes the

1     average maximum speed and $a\Phi_{ij'}/w$ again the proportion of decelerated ants. For the

2     symmetry restoring transition with $D^2 = 0$, Eq. (7) requires $a\gamma\phi/w < 1/3$, which implies

3     $V_{ij} > V_m[1-1/(3\gamma)] \approx 0.42 V_m$. However, according to the *empirical* speed-density relation by

4     Burd et al.[23], the maximum flow (the capacity) is only reached at the smaller speed

5     $V_{ij} = V_m[1-1/(n+1)] \approx 0.39 V_m$. Although the empirical value $n \approx 0.64$ was determined for

6     another ant species, the values for *L. niger* should be comparable. This shows that the

7     symmetry restoring transition occurs *before* the maximum flow is reached. The strict

8     inequality also implies capacity reserves and a limitation of the density-related speed

9     reduction. A marginal reduction in speed, however, would not be in favour of symmetrical

10    traffic because of benefits by using a single trail: First, a more concentrated trail provides a

11    better orientation guidance and a stronger arousal stimulus[24]. Second, a higher density of ants

12    enhances information exchange and supports a better group defence[25].

13    To conclude, this study demonstrates the surprising functionality of collisions among ants to

14    keep up the desired flow level by generation of symmetrical traffic. Pushing behaviour may

15    be considered as an optimal behaviour to maintain a high rate of food return to the nest. It

16    would not be required in most models of Ideal Free Distribution[26] (IDF), as they neglect

17    effects of inter-attraction. The balancing between cohesive and dispersive forces avoids a

18    dysfunctional degree of aggregation and supports an optimal accessibility of space at minimal

19    costs allowing an efficient construction, maintenance and use of infrastructures. This

20    mechanism appears to be generic in nature, in particular in group living organisms. For

21    example, inhibitory interactions at overcrowded building sites in termites allow a smooth

22    growth of the nest structure[27]. In the development of urban agglomerations they are important

23    to keep up the co-existence of cities[28] and in vehicle traffic, they determine the choice of

24    longer, less congested routes. The mechanism also suggests algorithms for the routing of data

25    traffic in networks.

1  Competing interests statement: The authors declare that they have no competing financial interests.

2  Correspondence and requests for materials should be addressed to dussutou@cict.fr

**Figure Legends**

**Figure 1** Experimental set-up. Five queenless colonies of *Lasius niger* (Hymenoptera, Formicidae) each containing 500 workers were used. During an experiment, ants had access from the nest to a source of sucrose (2mL of 1M solution) placed at the end of a diamond-shaped bridge. The solution was spread over a large surface so that all ants arriving at the end of the bridge could have access to the food. $j = 1$ and $j = 2$ indicate the choice points for branch selection. The whole set-up was surrounded by white curtains to prevent any bias in branch choice due to the use of visual cues. The colonies were starved for four days before each experiment. The traffic on the two-branch bridge was recorded by a video camera for one hour. Nestbound and outbound ants were then counted and aggregated over 1-minute intervals. Counting began as soon as the first ant climbed the bridge. Four different bridge widths *w* were used: 10.0, 6.0, 3.0 and 1.5mm. This set-up mimics many natural situations in which the geometry of the trails is influenced by physical constraints of the environment such as the diameter of underground galleries or of the branches in the vegetation along which the ants are moving.

**Figure 2 a-d** Average number of ants per minute crossing the two branches of the bridge within intervals of five minutes. The observed traffic volumes $2\phi(t)$ and flow dynamics agreed for the four different branch widths, i.e. the bottleneck for smaller branches was compensated for by usage of both branches. $N = 12$ experiments were carried out for each bridge width. Error bars indicate the standard deviation.

**Figure 3** Experimental results. **(a)** Experimental frequency distributions of the proportion $F_{1j}$ of ants using the right branch for different branch widths *w*. We considered traffic to be asymmetrical when more than 2/3 of the cumulated traffic of ants at the end of the experiment had used a single branch. All experiments have been pooled. **(b)** Experimental frequency distributions similar to (a), but for different

total numbers of ants crossing the bridge. All experiments on bridges of 1.5, 3.0 and 6.0 mm width have been pooled. **(c)** Proportion of outbound ants on the left and right branches of the bridge to the total number of ants that have passed the bridge at the end of the experiment. The results contradict opposite one-way flows on both branches. **(d)** Number of pushing events as a function of the total number of ant encounters on the initial part of each branch of the bridge. For each branch width, pushing events were evaluated at both choice points, for outbound and nestbound ants, during the first ten and last ten minutes of a random sample of two experiments characterised by symmetrical and two experiments characterised by asymmetrical traffic. This yielded a total of 2 branches x 4 experiments x 2 time intervals = 16 points per branch width. When the number of encounters was too low ($\leq 3$) the points were not taken into account in the regression. The slope of the linear regression is equal to 0.571 $\pm$ 0.057 (CI$_{95\%}$). This corresponds to the probability $\gamma$ of an ant to be pushed and redirected towards the other branch after encountering another ant coming from the opposite direction (see **e-g**).

**Figure 4 Top:** Analytical results for the parameter values $q = 1$, $k = 6$, $\nu = 1/40$ min$^{-1}$, and $a = 0.1$ mm.min, which have been adjusted to experimental measurements. The curves show the proportion $F_{ij}$ of the overall flow $\phi$ on each branch in the stationary state. **(a)** In the absence of pushing ($\gamma = 0$) the model predicts asymmetrical traffic above very low values of the overall flow of ants, independently of the traffic volume. **(b)** When the proportion of pushed ants is high ($\gamma = 0.6$), traffic is asymmetrical for moderate values of the overall flow of ants, but stable symmetrical traffic with $F_{ij}=0.5$ is expected above a critical value of traffic flow which is an increasing function of branch width $w$. The symmetry-restoring transition from asymmetrical to symmetrical traffic flow corresponds to an inverse pitchfork bifurcation. **Bottom:** Results of Monte-Carlo simulations. At time $t = 0$, the pheromone concentration and the number of ants on each branch are set to zero. Ants arrive at choice point $j$ at a rate $\phi_j(t) = \phi$. The probability of choosing the right or left branch at a choice point is governed by the choice function $F_{ij}$ [see Eq. (3)]. However, as soon as an ant has engaged on a branch, it may be pushed to the other branch with probability $\gamma$ by an oppositely moving ant coming from the second choice point. The pushed ant then continues its course on the alternative branch and lays a trail. For each value of $\gamma$, the simulations are averaged over 1000 runs. The graphs show the relative frequency of simulations in which a certain proportion of traffic was supported by the right branch. **(c)** Assuming a pushing probability equal to zero ($\gamma=0$), most simulations ended with asymmetrical traffic. However, when we used the experimental value $\gamma=0.6$, most simulations for narrow branches resulted in symmetrical traffic. **(d)** The proportion of simulations with $\gamma=0.6$ in which asymmetrical traffic emerged is a function of the total number of ants crossing the bridge, just as in our experiments (cf. Fig. 3c).

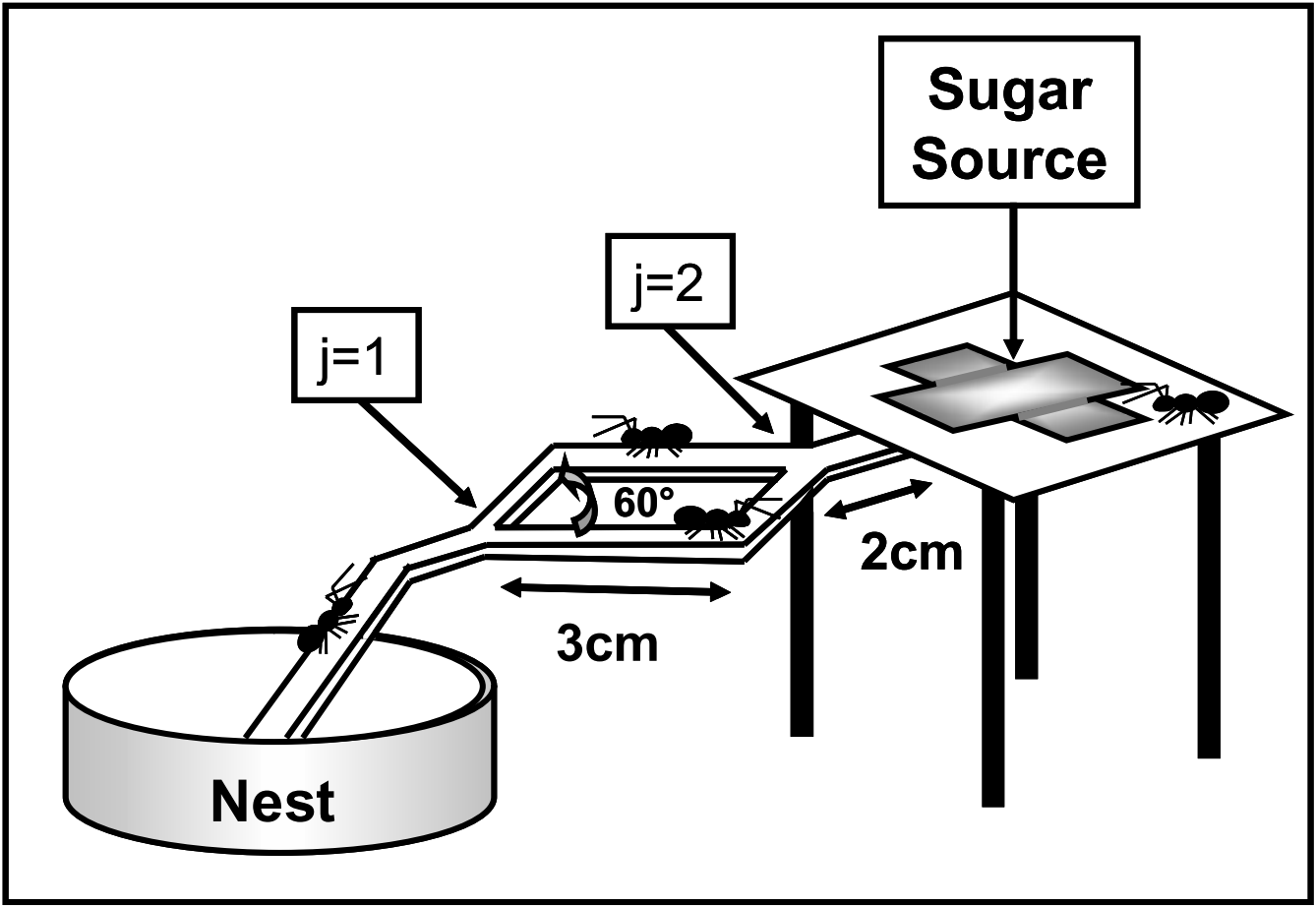

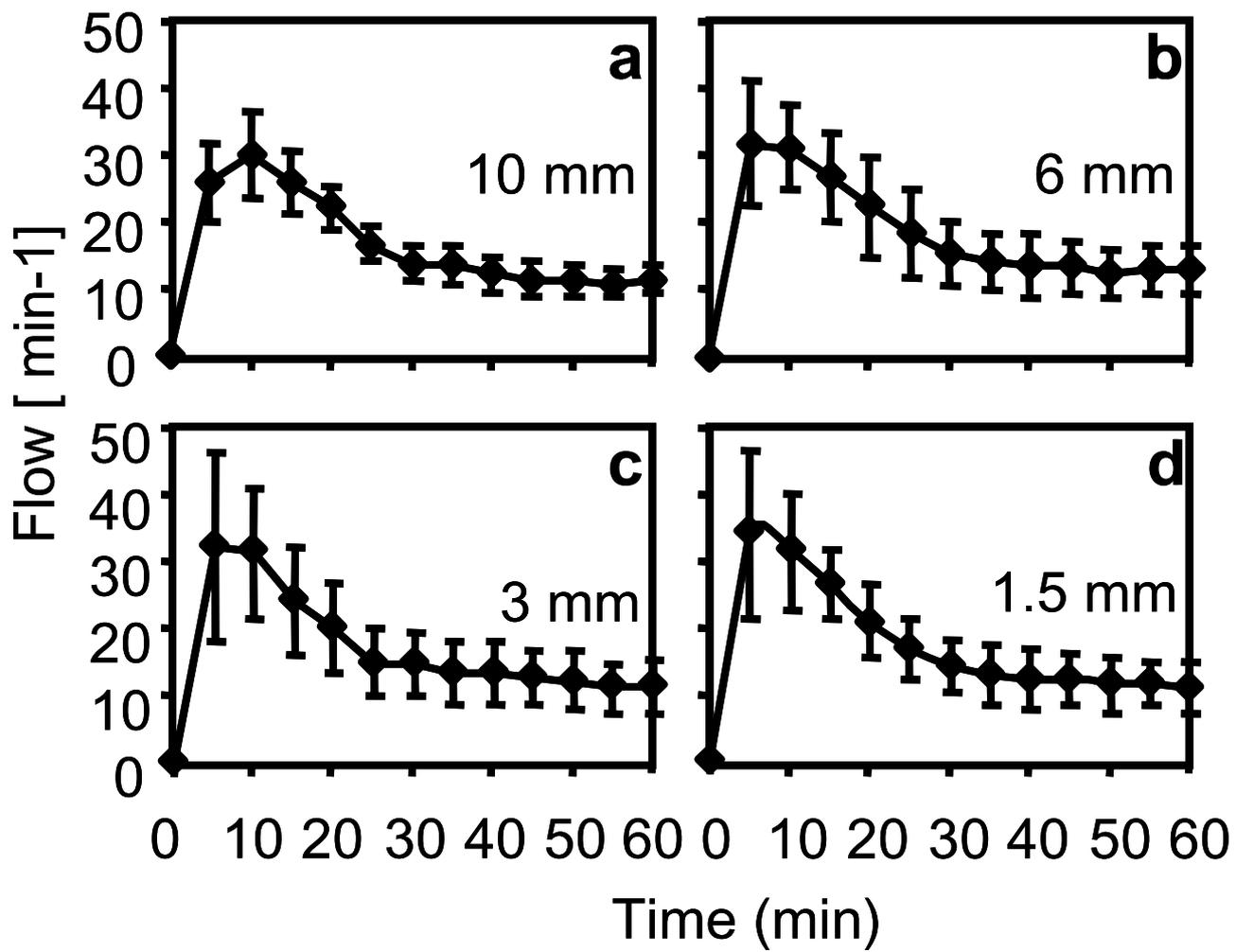

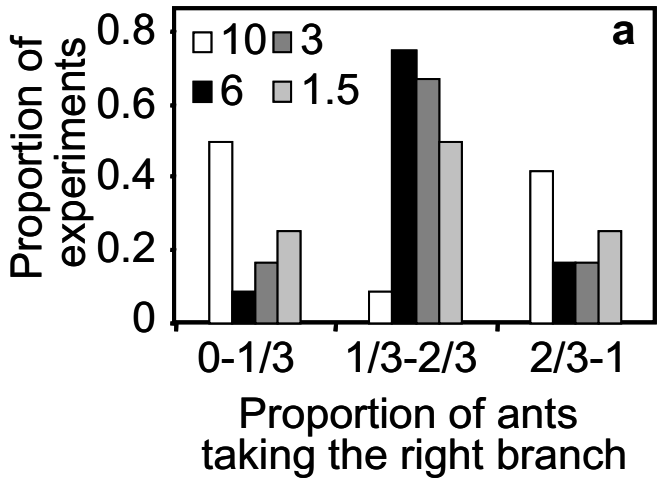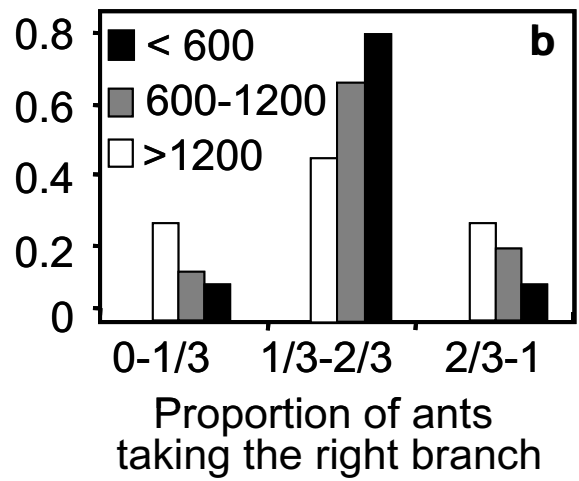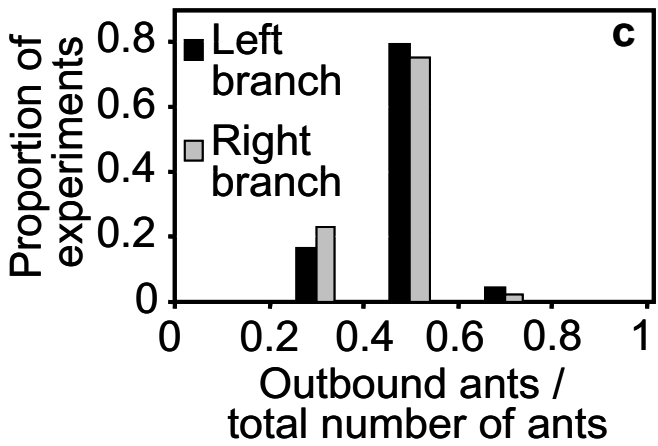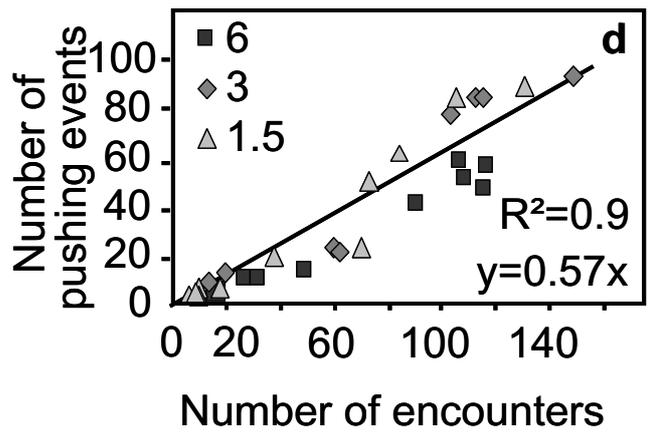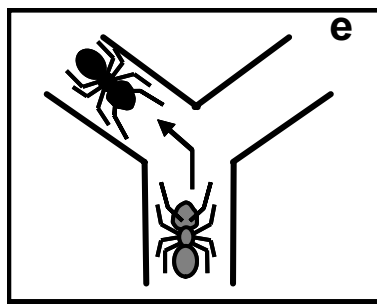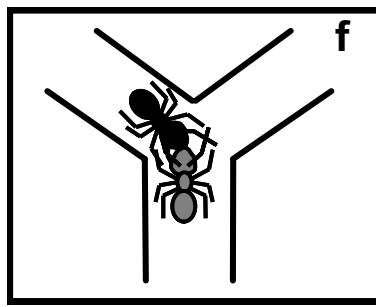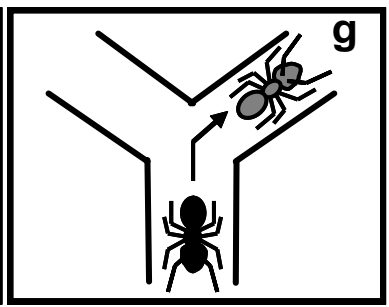

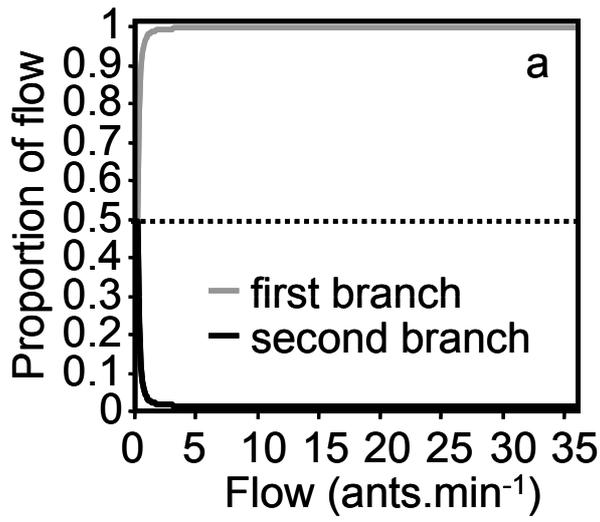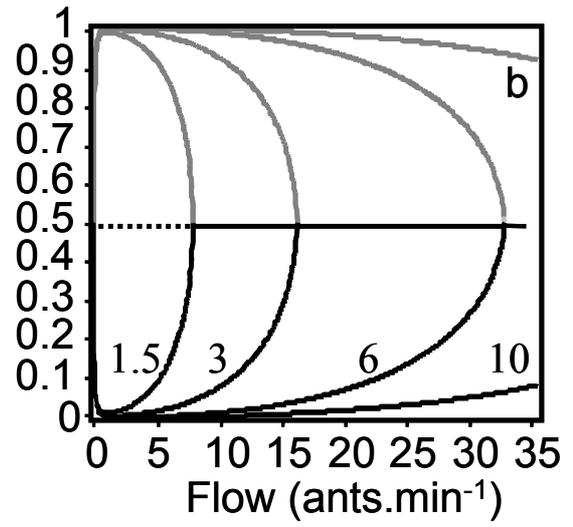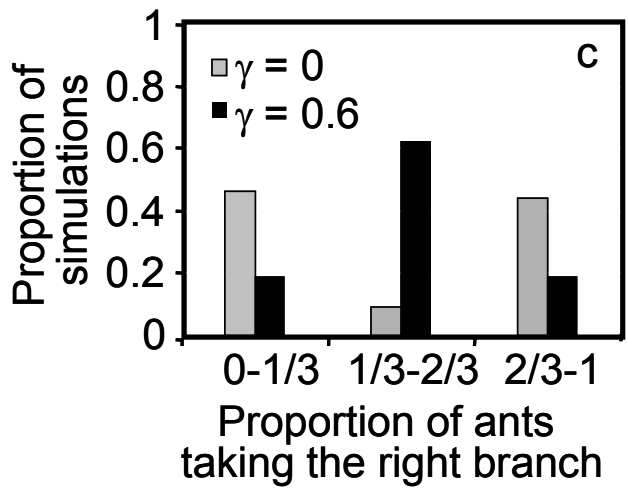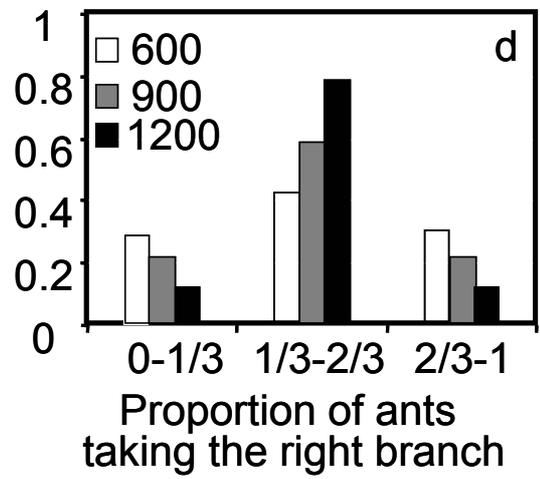